# Optimization of gridshell bar orientation using genetic algorithms

L. Bouhaya[a,*], O. Baverel[a, b], J-F. Caron[a]

[a] Université Paris-Est, Laboratoire Navier (UMR CNRS 8205), Ecole des Ponts ParisTech, 6-8 av Blaise Pascal, Champs-sur-Marne, 77455 Marne la Vallée Cedex 2, France

[b] Ecole Nationale Supérieure d'Architecture de Grenoble, 60, Avenue de Constantine, BP 2636, Grenoble Cedex 2, France

* Phone: (33) 1 64 15 37 91; fax: (33) 1 64 15 37 41; e-mail: lina.bouhaya@enpc.fr

**Abstract**

Gridshells are defined as structures that have the shape and rigidity of a double curvature shell but consist of a grid not a continuous surface. This study concerns those obtained by elastic deformation of an initially flat two-way grid. This paper presents a method to generate gridshells on an imposed shape under imposed boundary conditions. A numerical tool based on a geometrical method, the compass method, is developed. It is coupled with genetic algorithms to optimize the orientation of gridshell bars in order to minimize bar breakage during the construction phase. Examples of application are shown.

## 1. Introduction

Gridshells are often defined as structures that have the shape and rigidity of a double curvature shell but consist of a grid rather than a continuous surface. In this work, they are obtained by elastic deformation of an initially flat two-way grid. This reduces the grid's shear stiffness allowing large deformations. The deformed grid is then rigidified using a third direction of bars or panels. A gridshell thus has interesting structural potential and can respond to complex architectural requirements. A dozen gridshells have been constructed using the described technique, among them the Bundesgartenschau Mannheim Multihalle gridshell in Germany [1] and the Downland Museum in the United Kingdom [2]. In these gridshells, wood was chosen for its low density and relatively large limit strain. The Navier laboratory has recently proved the feasibility of such structures in composite materials. Two prototypes of gridshells in composite materials were constructed at the Ecole des Ponts ParisTech (figure 1) [3-4] and a 300 m$^2$ gridshell was built for the Solidays festival in June 2011 in Paris (figure 2).

Two methods have been used in gridshell formfinding, one experimental, the inversion method [5]; and one numerical, principally the dynamic relaxation method [6-7-8-9]. The inversion method is based on the assumption that the flexural stiffness of the grid elements is negligible. This method was used in the design of the Mannheim Bundesgartenschau gridshell in Germany, the first and largest timber gridshell ever built. The second method, the dynamic relaxation method, is a numerical tool that uses dynamic calculation to find the static equilibrium state of a mechanical system. The Downland museum gridshell in the United Kingdom and the two prototypes of gridshell in composite materials at the Ecole des Ponts ParisTech were designed using this method. Both techniques, the inversion method and the dynamic relaxation method, lead to a deformed grid through calculation. The shape obtained is close to the one proposed by the architect but is difficult to control. These methods are described in the first part of the paper.

In this paper, the first aim is to focus on the generation of gridshells on an imposed form and under imposed boundary conditions. By definition, mapping a gridshell on a form is equivalent to drawing two-way parallel and equidistant curved axes, i.e. parallelograms, on the surface. Mathematically speaking, those nets are called Tchebychev nets [10-11]. The problem is very similar to that encountered with fabric draping. Among the methods used for composite forming, a geometrical method, the fishnet algorithm, was introduced by Van Der Waëen [12]. Other mechanical approaches are described by Boisse in [13-14]. A method for mapping a two-way elastic grid using finite element simulations was inspired from those approaches and was introduced previously in [15].

A geometrical method that allows the generation of a Tchebychev net on an imposed surface, called the compass method, was introduced in IL10 Gitterschalen [5]. This method is similar to the basic version of the fishnet algorithm. A numerical tool based on the compass method is developed. Its main steps are described in the second part of the paper.

An elastic gridshell is subjected to stresses due to its construction procedure. The second aim of this work is therefore to optimize the gridshell's bar orientations in order to minimize bar breakage during the construction phase. This is done by coupling the compass method with genetic algorithms [16]. The algorithm optimizes the gridshell by minimizing bar curvatures. We suppose that the bracing of the shell could be made with a third direction of bars that have reduced section or with panels as in the Savill Building in England [17]. The proposed algorithm is described in the third part of the paper and several examples of applications are shown in the last part.

## 2. Gridshell and formfinding

As mentioned before, gridshell formfinding has been effected historically essentially by means of two methods, inversion method and dynamic relaxation.

The inversion method is an experimental technique whose concept was first introduced by Hooke in 1675 [18]. Hooke described the relationship between a hanging chain, which forms a catenary in tension under its own weight, and an arch in compression. Frei Otto applied this concept to gridshell formfinding [5]. This method is based on the assumption that the flexural stiffness in the grid elements is negligible. The inversion method can be summarized as follows: First, a net with square meshes is constructed. Then, a set of boundary conditions is chosen and the net is suspended. The geometry of the net is raised by photogrammetry. Finally the positions of nodes are recomputed with the force density method so that the structure is in equilibrium under actual weight.

Figure 3 shows a suspended form made by the German Institute for lightweight structures *für leichte Flächentragwerke* and by turning the corresponding gridshell. This method was used in the conception of the Mannheim Bundesgartenschau gridshell. Figure 4 shows the inversion form of the Mannheim gridshell.

The second technique used for gridhsell formfinding is a numerical one based especially on the dynamic relaxation method. The concept of dynamic relaxation was introduced by Day [6] and was applied for the formfinding of tension structures by Barnes [7-8-9]. The dynamic relaxation method uses a dynamic calculation to find the static equilibrium state of a mechanical system. It provides the determination of the static equilibrium state of a structure by seeking the maximum kinetic energy which corresponds to the minimum potential energy for a conservative system. This procedure allows the movement of the structure to be drawn. When a peak total kinetic energy is detected, all velocities are reset. After that, the procedure is repeated and the structure is free to oscillate again until the next maximum kinetic energy. And so on, until the kinetic energy of all modes of vibration has been dissipated and equilibrium has been reached.

This method was adapted and used for the formfinding of slender prestressed structures in composite materials [4]. First a flat two way direction grid is chosen. Then it is pushed upwards at every connecting node. The boundary conditions remain in the initial plane of the grid. Finally, a third direction beam is set in place to rigidify the structure. This method was used to design the two prototypes of composite materials gridshells that were built on the site of the Ecole des Ponts ParisTech. These gridshells are made of pultruded glass fibre reinforced tubes with a 42mm in diameter and 3.5mm thick. The roofs are made of polyester canvas-coated PVC, with the help of Ferrari SA, Esmery Caron and Abaca consulting engineers. This method was also applied to the design of reciprocal frame systems called nexorades [19].

In both methods, the final shape is the result of an irreversible and iterative process. Hence the grid obtained is a result of calculation and cannot be controlled from the beginning. The aim of the following part is to be able to map a grid on an imposed form given by an architect.

## 3. Mapping a gridshell by use of the compass method

The compass method, described in IL10 Gitterschalen [5], is a geometric method that allows a network of parallelograms to be created on any surface by means, as its name suggests, of a compass. This type of grid is also known as a Tchebychev network [10]. In this section, the steps used to generate a gridshell on a surface using the compass method are described. Those steps are summarized by the flowchart in figure 5. They have been adjusted so that they can be coupled with the genetic algorithms described in the next section in order to minimize the curvatures and resultant levels of stress in the bars.

Let us consider the process of mapping a gridshell on a surface characterized by the Cartesian equation F (x, y, z) = 0 by using the compass method. The first step consists in drawing two arbitrary curved axes ($D_1$) and ($D_2$) that intersect at a single point A ($x_A$, $y_A$, $z_A$) as shown in figure 6.

In order to draw the guidelines, let $\alpha_1$ be the angle between the director ($D_1$) and the direction of the x-axis at point A. The position of point A ($x_A$, $y_A$, $z_A$), the angle $\alpha_1$ and the mesh width w are known, the first point $M_1$ of the guideline ($D_1$) can be determined (Figure 7). It represents the intersection of surface S, plan (P) passing through A which is at angle $\alpha_1$ with (xOz) plan and the sphere with A as the centre and w as the radius. In the same manner, the other points $M_i$ of the director ($D_1$) can be determined given the position of $M_{i-1}$, the mesh width w and the angle $\gamma_{i-1}$ between ($M_{i-1}M_i$) and ($M_{i-2}M_{i-1}$) (Figure 8). $M_i$ results from the intersection of surface S, the plan ($P_{i-1}$) passing by $M_{i-1}$ and forming angle $\theta_{i-1}$ with the plan (xOz) and the sphere with $M_{i-1}$ as the centre and w as the radius.

By proceeding in this way, the two axes ($D_1$) and ($D_2$) can be drawn on the surface. Thus,

all the points $M_i$ and $N_j$ can be plotted respectively, on guidelines $(D_1)$ and $(D_2)$. Now, let us consider a part of the area limited by a part of axis $(D_1)$ to the right of point A and that of $(D_2)$ to the bottom of point A. Starting with point A which is the intersection of the axes $(D_1)$ and $(D_2)$, two neighbouring nodes $M_1$ and $N_1$ have been defined. The fourth node P(1,1) is the intersection of the surface and the two spheres drawn around each of the neighbouring points with radii equal to the mesh width w (figure 9). Then, gradually, new points are determined in the same way (figure 10) and the net nodes are connected rectilinearly (figure 11). Similarly, the other three parts of the surface can be meshed.

In order to apply this method on a surface with the Cartesian equation, the following parameters, shown in Figure 12, must be remembered:

- w: the mesh width.

- A $(x_A, y_A, z_A)$: the point of intersection of two guidelines $(D_1)$ and $(D_2)$, considered as the point of mesh initiation.

- $\alpha_1$: the angle at point A of axis $(D_1)$ with the direction of x-axis (at the right of A).

- $\alpha_2$: the angle at point A of axis $(D_1)$ with the direction of x-axis (at the left of A).

- $\beta_1$: the angle at point A of axis $(D_2)$ with the direction of x-axis (at the top of A).

- $\beta_2$: the angle at point A of axis $(D_2)$ with the direction of x-axis (at the bottom of A).

- $\gamma_{1i}, \gamma_{2j}, \varepsilon_{1k}$ and $\varepsilon_{2l}$: changes of angles in the two guidelines $(D_1)$ and $(D_2)$.

Figure 13 shows three examples of surfaces that have been mapped using the compass method. The mesh was made on the bases of two orthogonal guidelines. The surfaces are, going from the top to the bottom: a positive double curvature surface, i.e. the ellipsoid, a negative double curvature surface, i.e. the first Scherk surface, and a surface with a positive and negative double curvature, i.e. the torus.

In this way, a gridshell can be mapped on imposed forms using the compass method. In order not to break the bar during the erection stage and once the gridshell is set up, the bending stress and curvature of the bar should be as low as possible. A specific algorithm has been developed in order to find the optimum path of each bar on a proposed surface which also respects the construction of the grid.

## 4. Gridshell optimization: compass method coupled with genetic algorithms

In order to minimize the stresses in the bars due to the construction of the gridshell mapped with the compass method, an optimization tool is introduced in this section. The aim is to find the network of bars having the lowest bending stresses. This mesh can be obtained by minimizing the maximum curvature in the bars of the entire gridshell. In this case, the cost function is difficult to express. This function is not a minimization of the global strain energy but a minimization of the maximum curvature. Hence, a stochastic algorithm has been investigated.

Genetic algorithms (GAs) are stochastic optimization algorithms based on genetic mechanisms and natural evolution, such as selection, crossover and mutation. They were developed by John Holland in 1970 [20]. Those methods have proven effective in structural optimization problems [21-22-23]. Genetic algorithms or evolutionary algorithms have been chosen due to the versatility of the method [21]. The proposed algorithm is described in the following section. It can be summarized by the flowchart in figure 14.

### 3.1 Optimization Algorithm proposed

The proposed algorithm based on genetic algorithms is summarized in the following steps:

1. Random generation of a population of *npop* individuals:

To apply genetic algorithms to the compass method, the mesh must be represented in the form of a chromosome. A chromosome is defined as a potential solution to the problem. It is chosen as a group of genes that define the mesh. These genes are the point A which is the intersection point of the two guidelines and all the angles needed to define the two guidelines. Each time a chromosome is chosen, it must be verified that the surface is mapped completely and without overlapping.

2. Evaluation of each individual:

The "fitness function" is the evaluation function of the problem. It is the function that evaluates each chromosome and determines whether it can survive or not. We are interested in this work to minimize the stresses, and therefore the curvatures in the gridshell bars. The aim of this algorithm is to map a surface with the least curved bars. Thus, we propose a fitness function which is equal to the maximum curvature, obtained as a result of mapping the surface by the chromosome using the compass method. This fitness function could also be the mean curvature, or the mean of a percentage of the maximum curvature values.

To calculate the curvature, let us consider a curved line represented by three points $M_{i-1}$, $M_i$ and $M_{i+1}$ as shown in figure 15. The radius of curvature R and curvature C can be calculated as follows:

$$R_i = \frac{\left\| \overrightarrow{M_{i+1} M_{i-1}} \right\|}{2.\sin \alpha} \quad \text{and} \quad C_i = \frac{1}{R_i} \quad (1)$$

For each chromosome, a mesh can be generated on the surface considered, and the curvatures computed at each grid point by equation 1. The maximum value of the curvatures is taken and is considered as the fitness function of the chromosome in question.

3. Cycle of evolution:

    (A) Selection of a set of parents:

    Two chromosomes called "parents" are chosen randomly from the population in order to create two chromosomes called "children". Several methods of parent selection can be used. One called "Tournament paring" is used in this work. It consists of selecting a set of chromosomes *ncan* of the population each time we want to select an individual "parent". The chromosome with the best fitness in the group is taken as parent.

    (B) Crossover with crossover probability *pc*:

    In this step, two child chromosomes are created from two parent chromosomes selected in the previous step. Uniform crossover is used in this work. A random number between 0 and 1 is generated for a pair of genes of parents $P_1$ and $P_2$. If this number is lower than the probability of crossover *pc* chosen, the considered genes are not changed. However, if this number is greater than *pc*, the genes are exchanged between the chromosomes of the two parents.

    (C) Mutation with a mutation probability *pm*:

    A mutation probability *pm* between 0 and 1 is initially fixed. For each child gene, a random number is generated. If this number is lower than the mutation probability, the gene is mutated to a random number chosen from the interval in which the gene is defined.

4. Step 3 is repeated until the new population contains *npop* individuals. Each time, the two child chromosomes are classified in the population and two individuals with the greatest fitness functions are eliminated.

5. Iterate from step 2 until the convergence of the algorithm:

When the difference between the best values of the fitness function for two successive generations is small enough, no improvement was made in passing from one generation to another. In this case, the best fitness value of the current generation is chosen as the solution for the problem. The convergence problem can be written as:

$$\left|\frac{f_i - f_{i-1}}{f_i}\right| < \varepsilon \quad (2)$$

## 5. Examples of application

The optimization procedure described in the previous section was applied on several surfaces. In this section, applications of this method on a surface with Gaussian positive curvature i.e. a hemisphere, a surface with Gaussian positive and negative curvature i.e. a surface with a sinusoidal equation for its Cartesian equation and a surface with Gaussian negative curvature i.e. a hyperbolic paraboloid, are shown.

### 5.1 Hemisphere (Gaussian positive curvature)

The first surface considered is a hemisphere with O (0,0,0) as its center and 10 units as its radius. The aim is to optimize the mesh obtained by the compass method applied with a mesh width of two units. An initial population of 100 chromosomes is randomly chosen. Each chromosome is composed of 50 genes. The optimization procedure described in Section 3.1 is applied with a mutation probability of 0.01 and a crossover probability of 0.5. Point A, the point of intersection of two guidelines is chosen randomly in the area limited by x = -2, x = 2, y = -2 and y = 2 axes.

Once the convergence is checked, the graph in the figure 16 can be drawn. It shows the mean and the best fitness of each generation. In order to compare the solution obtained by the proposed algorithm, two calculations were made, using intuitive trial and error and the drawing of lots respectively. The first one consists in varying the angle between the two guidelines and searching for the solution that gives the minimum fitness function. It corresponds to a 70° angle between the two guidelines in the case of hemisphere. This solution is represented in figure 16 by the green line with y=0.178 equation. The second calculus consists in drawing 10000 chromosome lots. The number of chromosomes was chosen in order of proposed algorithm magnitude. The black line of the graph with y=0.187 equation represents the minimum value of the maximum curvature that can be reached while drawing 10000 chromosome lots.

The average fitness function of the 100 chromosomes of the first generation was of 0.2262 $m^{-1}$, and the fitness function of the best chromosome of the first generation was 0.196 $m^{-1}$. After 51 generations, convergence is reached. The final generation has an average fitness function of about 0.1718 $m^{-1}$. Thus, starting from 100 chromosomes, the result obtained is improved by 8% compared to the drawing of lots and 3.5% compared to the trial and error method. The mesh of the surface by the chromosome taken is given in figure 17.

A study was done by varying the number of chromosomes of the initial population of 50 to 900 chromosomes. It was noticeable that the greater the number of chromosomes, the smaller the fitness function. Thus, starting with 900 chromosomes, the result obtained is 16.5% better than the drawing of lots and 12% better than the trial and error calculus. It was remarkable also to see the low influence of the position of point A, the point of intersection of the two guidelines. This is due to the fact that the surface considered, i.e. the sphere, is symmetric.

## 5.2 Sinusoidal surface (Gaussian positive and negative curvature)

The second surface considered is a surface with double curvature presenting no symmetry. This surface has the sinusoidal Cartesian equation z = 0.05. x. sin (x) + sin (y) for 0< x<10 and 0<y< 4. It is mapped with the compass method with a mesh width equal to one unit. The optimization procedure described above was applied on an initial population of 100 chromosomes, a mutation probability of 0.01 and a crossover probability of 0.5.

Figure 18 shows the mean and best fitness of each generation. The black line with the y=0.405 equation represents the minimum value of the maximum curvature that can be reached while drawing of 10000 chromosome lots. The green line y=0.467 represents the minimum of $C_{max}$ that can be obtained by a trial and error calculus.

The result obtained after convergence has an average fitness function of about 0.378 $m^{-1}$. Thus, starting with 100 chromosomes, the result obtained is improved by 7%, compared to the drawing of lots and 19% compared to the trial and error method. The surface mesh by the chromosome obtained is given in figure 19.

As in the case of the hemisphere, a sensitivity analysis of the number of chromosomes in the initial population was conducted by varying from 100 to 700 chromosomes. Among the calculations made, the calculation made on 700 chromosomes represents the calculation for which the maximum curvature of the grid is the smallest. This value is 0.36 $m^{-1}$. Thus, starting from 700 chromosomes, the result is better by 11.5% than the drawing of lots and by 23% than the trial and error.

## 5.3 Hyperbolic paraboloid (Gaussian negative curvature)

The third surface considered is a hyperbolic paraboloid with the Cartesian z = $x^2-y^2$ equation with -1< x< 1 and -1<y <1. The mesh is made using the compass method with a mesh width of 0.3 units and A (0,0,0).

A hyperbolic paraboloid can be defined as a ruled surface; it could be generated by lines. On the surface considered, the least curvature lines can be drawn at 45° to the surface parabolas. The distance between two points taken on different parts of the two lines of the least curvature is not constant. The compass method, as described in paragraph 2, leads to a regular mesh whose width mesh is constant. Hence, if we consider mapping the surface in the direction of the lines of least curvature angles (45° and 135°), the maximum of curvature is equal to 1.39 m$^{-1}$.

The graph in figure 20 represents the mean and the best fitness of each generation. The calculation is performed on an initial generation of 700 chromosomes, a crossover probability of 0.5 and a mutation probability of 0.01. The chromosome number was chosen following the results of the sensitivity study on the number of chromosomes over the fitness function done in the two above-mentioned examples.

The black line with the y = 1.24 equation represents the minimum value of the maximum curvature obtained by drawing of 10000 chromosome lots. The green line with the y = 1.85 equation represents the value of Cmax that can be obtained by trial and error method, while the orange line represents the maximum curvature obtained by mapping of 45° and 135°.

The result obtained is 33% better than the best one obtained by randomly drawing 10 000 chromosomes, 55% better than the one generated by the trial and error method, and 40.3% better than the mesh of 45° and 135°. The hyperbolic paraboloid mesh by the chromosome obtained is given in figure 21.

## 6. Conclusions

A method to find and to optimize the orientations of the bars of a gridshell is proposed in this paper. It is based on the compass method coupled with genetic algorithms. It allows a gridshell to be mapped on an imposed shape and imposed boundary conditions proposed by an architect. It also allows us to optimize the orientations of the bars in order to minimize the curvatures and therefore the stresses in the structure during the construction and life time of the gridshell. Examples of application of this methodology are shown on a hemisphere and two surfaces with double curvatures. In the case of a simple symmetric shape, such as the hemisphere, the trial and error calculus can lead to results which are not far from the optimal solution. When the geometry becomes more complex, the calculation by trial and error or that of drawing lots is not sufficient to predict the direction of the bars that give low curvatures. The proposed algorithm allows us to minimize the maximum curvature on complex surfaces. For instance, the result obtained according to this algorithm is 33% better than a drawing of lots of 10 000 chromosomes. Optimization is done on the fitness function considered in this work as the maximum curvature in the bars of the resulting mesh. The versatility of the genetic algorithms allows for a choice of a different fitness function according to the demands of the designer.

Figure captions

Fig. 1. First prototypes of composite materials gridshells at Ecole des Ponts ParisTech

Fig. 2. Gridshell in composite materials for Solidays festival in Paris

Fig. 3. Suspended model of IL and, by inversion, the corresponding gridshell

Fig. 4. The suspended model of the Mannheim gridshell

Fig. 5. Compass method flowchart

Fig. 6. Illustration of the two guidelines on the surface

Fig. 7. Determination of point M1

Fig. 8. Determination of point Mi

Fig. 9. Determination of point P(1,1)

Fig. 10. Determination of point P(i,j)

Fig. 11. Surface mapping

Fig. 12. Nomenclature

Fig. 13. Mapping a) an ellipsoid, b) the first Scherk surface and c) a torus using the compass method

Fig. 14. Genetic algorithm diagram

Fig. 15. Curvature between three points

Fig. 16. Representation of the fitness function of each generation in the hemisphere

Fig. 17. The hemisphere mapped by the final chromosome

Fig. 18. Representation of the fitness function of each generation on the surface considered

Fig. 19. The surface mapped by the final chromosome

Fig. 20. Representation of the fitness function of each generation of the hyperbolic paraboloid

Fig. 21. The hyperbolic paraboloid mapped by the final chromosome

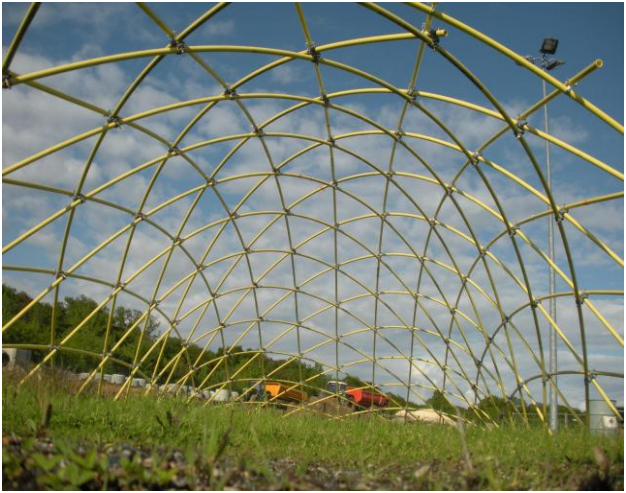 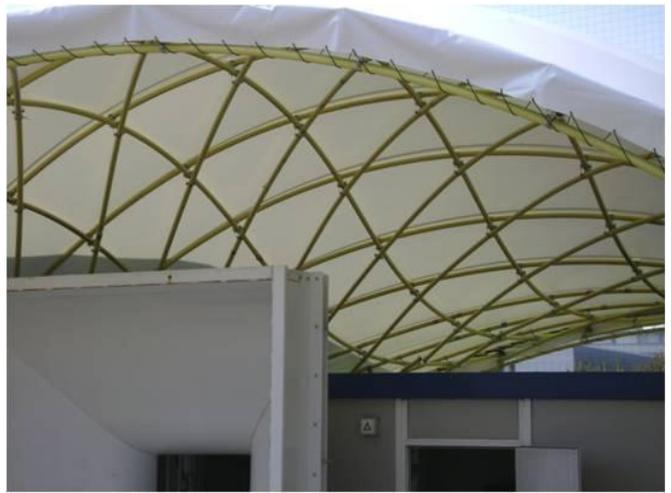

Fig.1. First prototypes of composite materials gridshells at Ecole des Ponts ParisTech

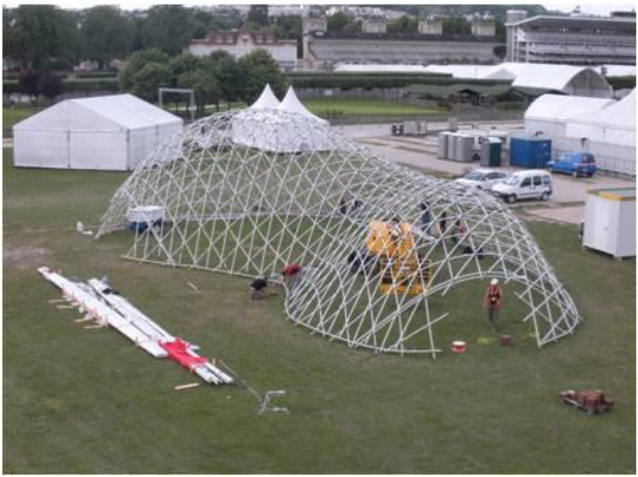 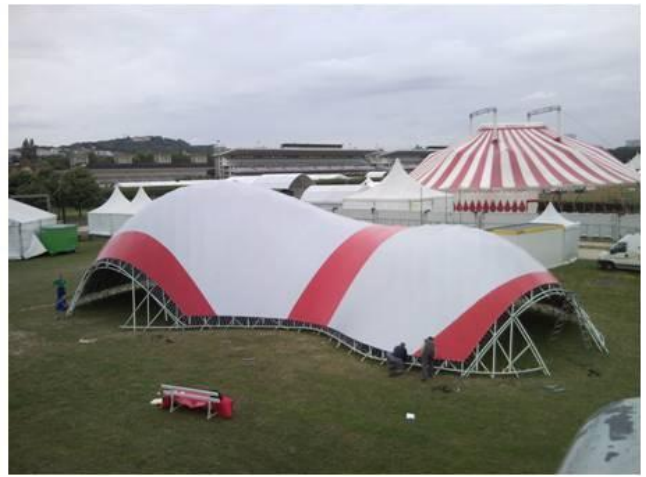

Fig. 2. Gridshell in composite materials for Solidays festival in Paris

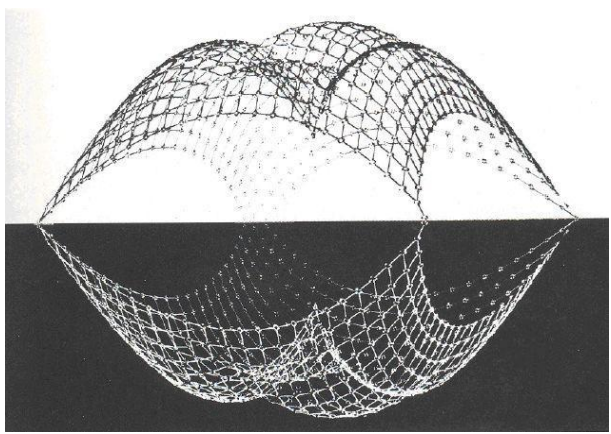

Fig.3. Suspended model of IL and, by inversion, the corresponding gridshell,

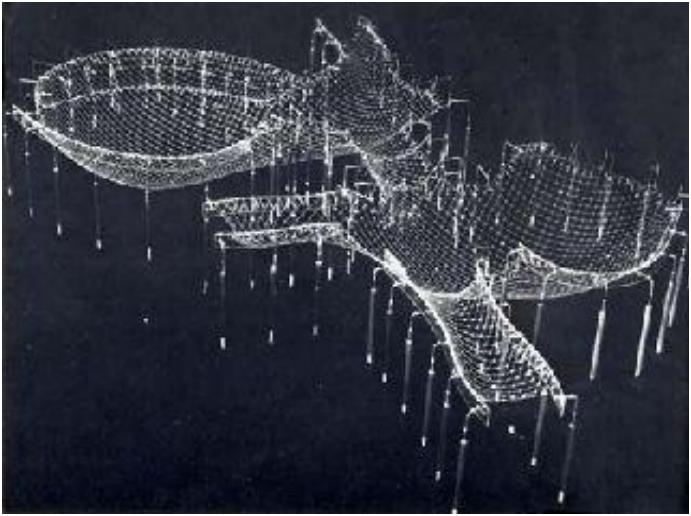

Fig. 4. The suspended model of the Mannheim gridshell

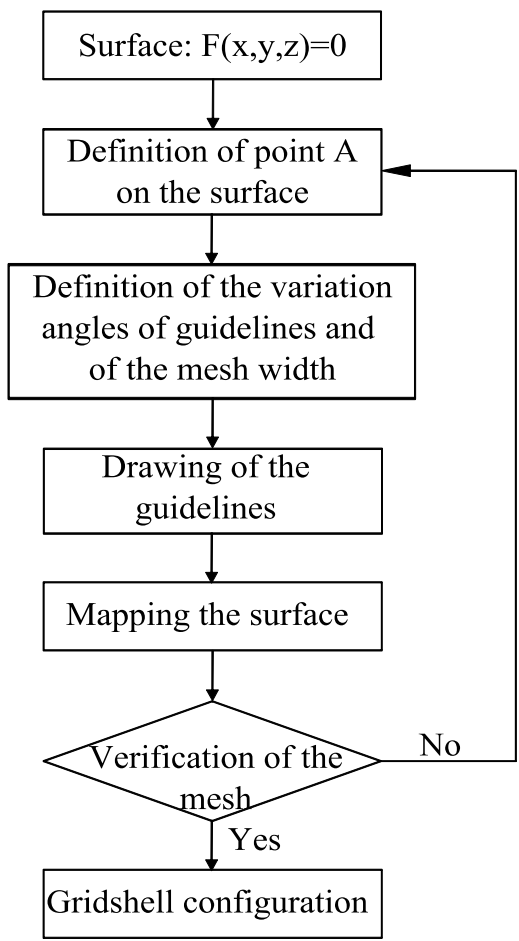

Fig. 5. Compass method flowchart

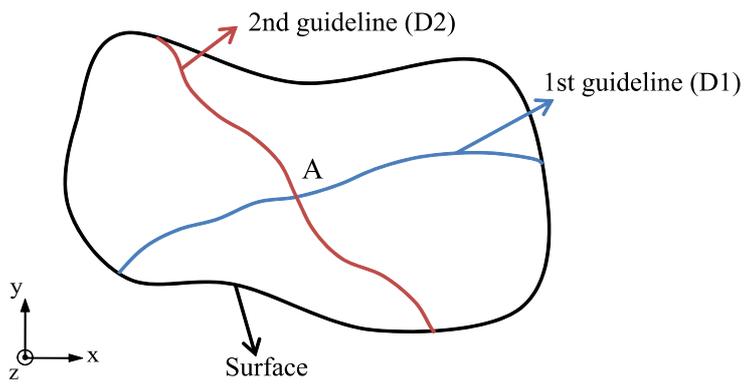

Fig. 6. Illustration of the two guidelines on the surface

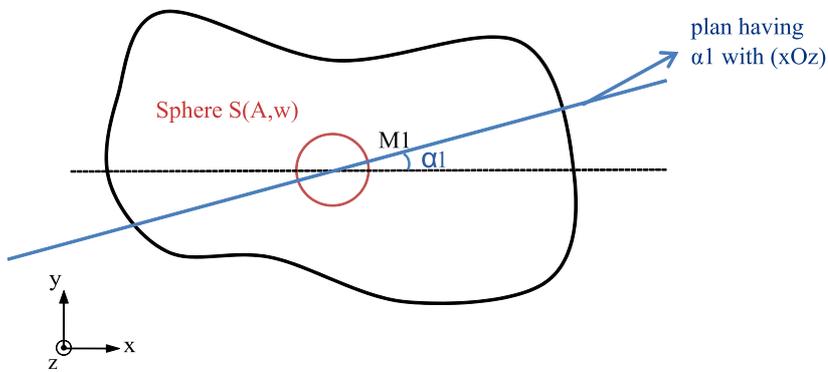

Fig. 7. Determination of point $M_1$

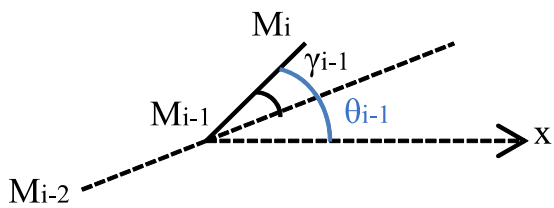

Fig. 8. Determination of point $M_i$

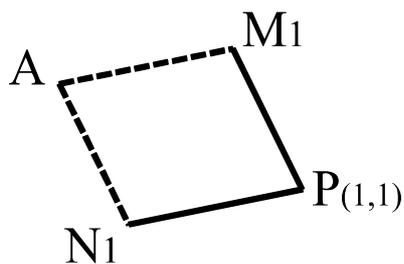

Fig. 9. Determination of point $P(1,1)$

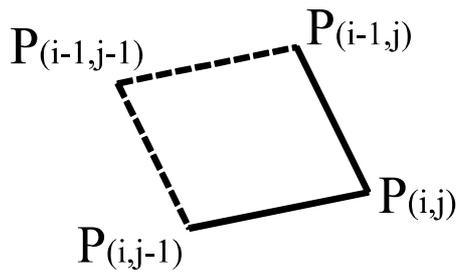

Fig. 10. Determination of point P(i,j)

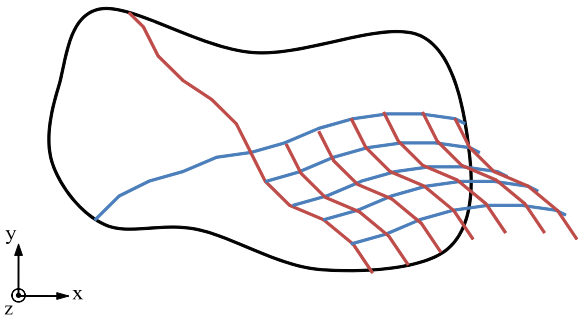

Fig. 11. Surface mapping

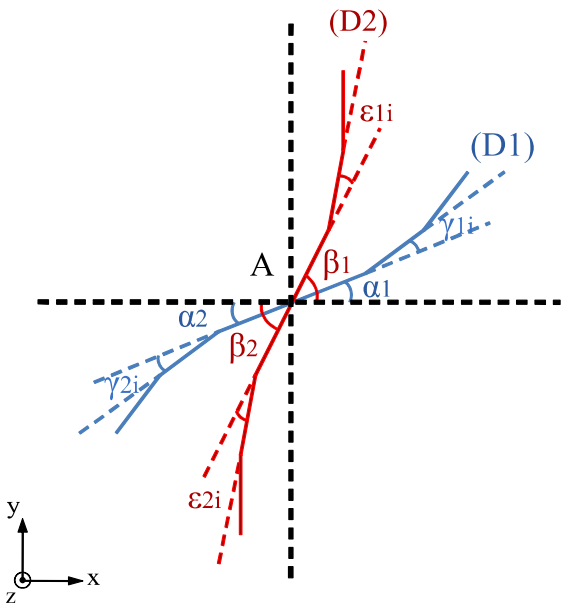

Fig. 12. Nomenclature

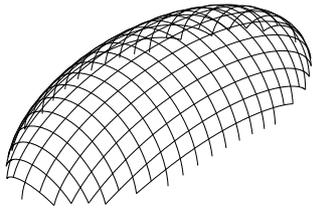 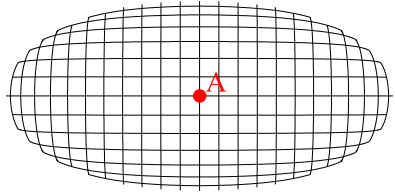

a)

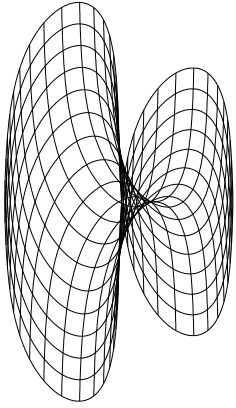 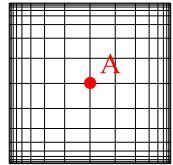

b)

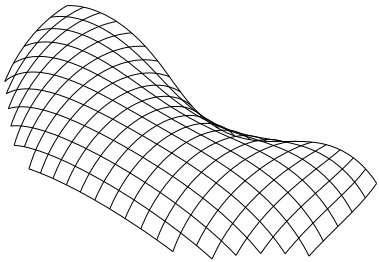 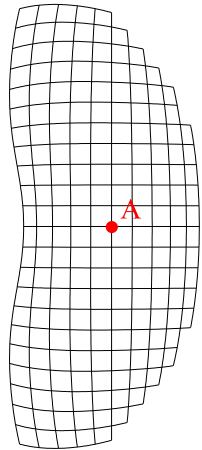

c)

Fig.13. Mapping a) an ellipsoid, b) the first Scherk surface and c) a torus using the compass method

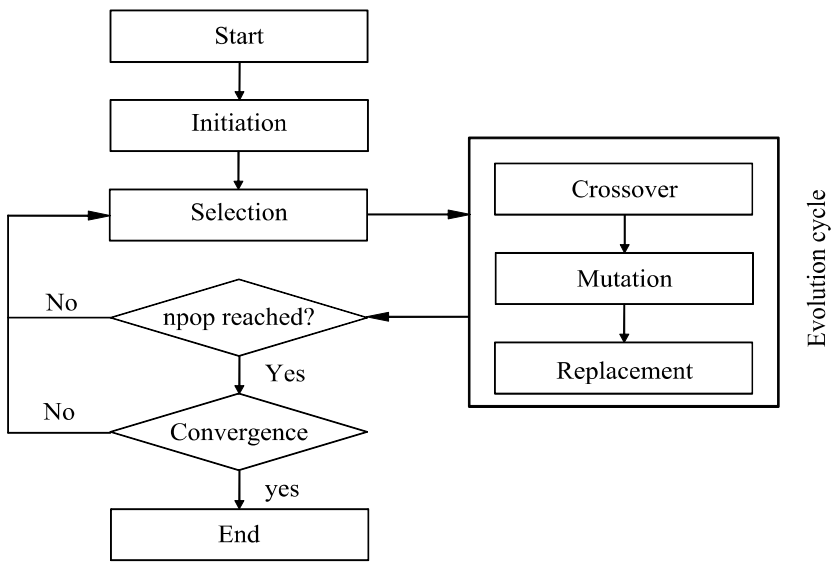

Fig. 14. Genetic algorithm diagram

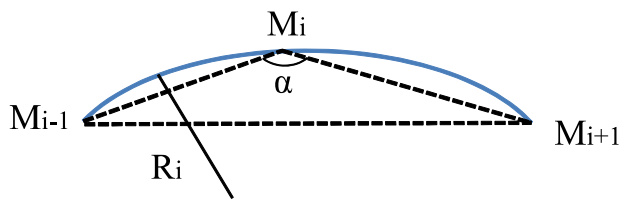

Fig. 15. Curvature between three points

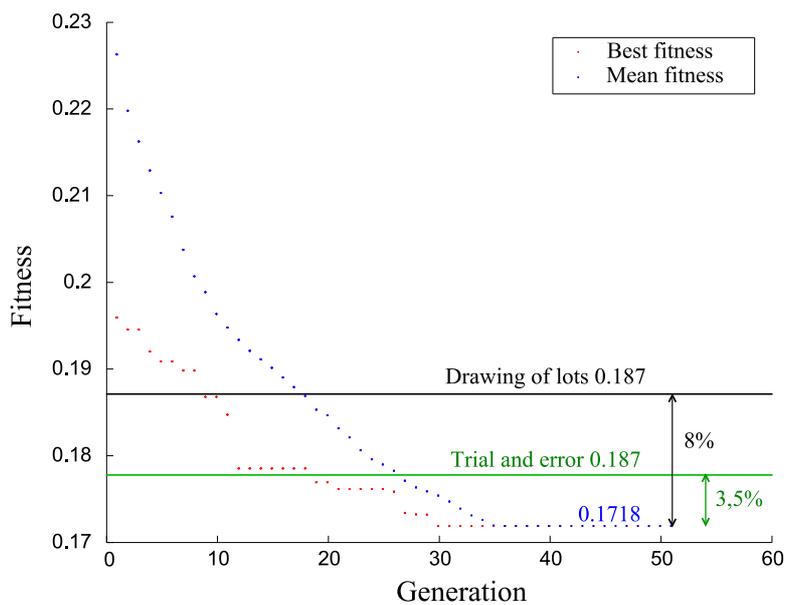

Fig. 16. Representation of the fitness function of each generation in the hemisphere

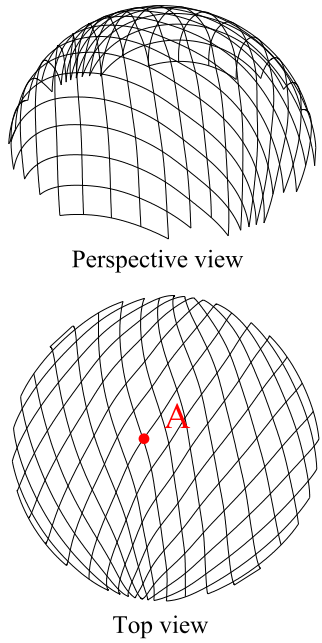

Perspective view

Top view

Fig. 17. The hemisphere mapped by the final chromosome

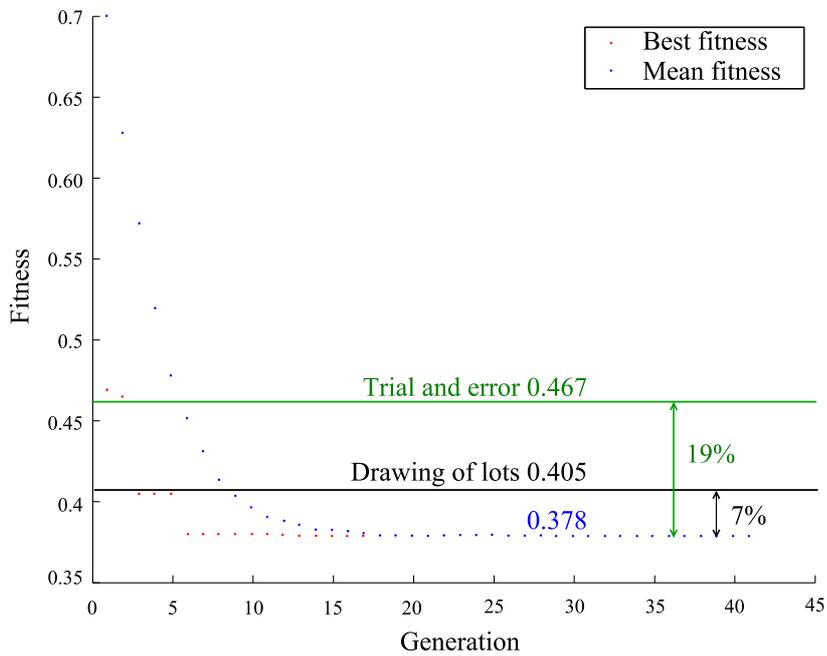

Fig. 18. Representation of the fitness function of each generation on the surface considered

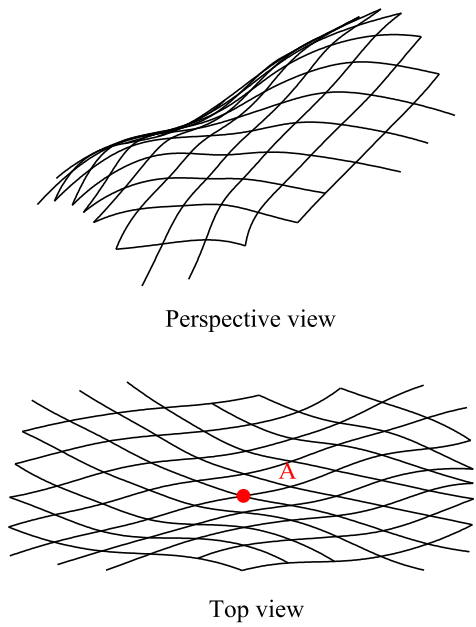

Fig. 19. The surface mapped by the final chromosome

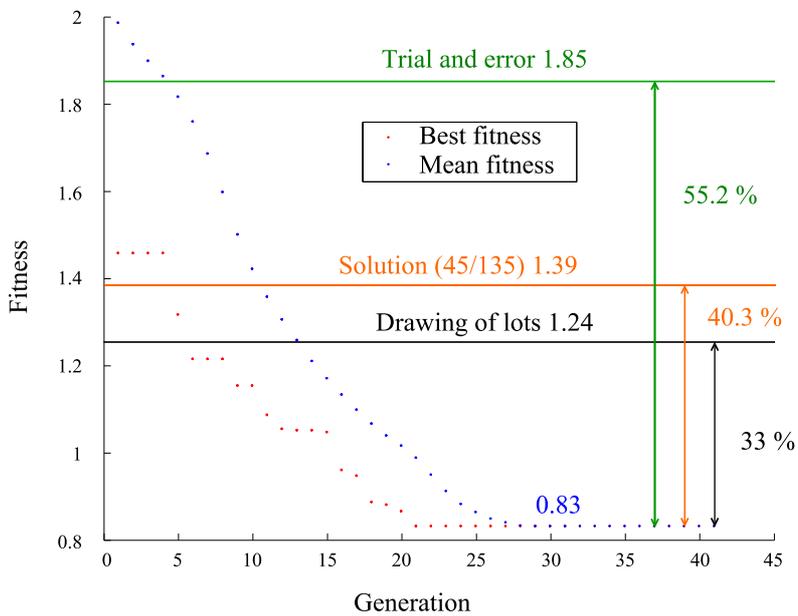

Fig. 20. Representation of the fitness function of each generation of the hyperbolic paraboloid

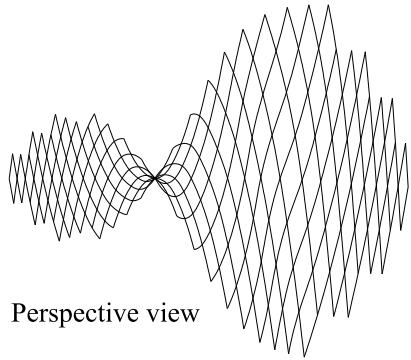

Perspective view

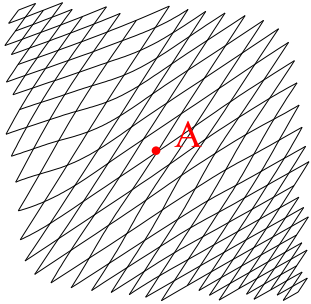

Top view

Fig. 21. The hyperbolic paraboloid mapped by the final chromosome